\documentclass[twocolumn]{aa} 
\usepackage{graphicx}
\begin{document}

  \title{An estimate of the time variation of the abundance gradient from 
   planetary nebulae\thanks{Based on observations made at the European Southern 
                 Observatory (Chile) and Laborat\'orio Nacional de Astrof\'\i sica 
                 (Brazil)}}

   \subtitle{II. Comparison with open clusters, cepheids and young objects}

   \author{W. J. Maciel,
          L. G. Lago,
          \and
          R. D. D. Costa
}

   \offprints{W. J. Maciel}

   \institute{Instituto de Astronomia, Geof\'\i sica e Ci\^encias
                 Atmosf\'ericas (IAG), Universidade de S\~ao Paulo - 
                 Rua do Mat\~ao 1226; 05508-900, S\~ao Paulo SP; Brazil\\
                 \email{maciel@astro.iag.usp.br, leonardo@astro.iag.usp.br, 
                 \\roberto@astro.iag.usp.br}
                         }

   \date{Received ; accepted }

   \abstract{The temporal behaviour of the radial abundance gradients 
   has important consequences for models of the chemical evolution of 
   the Galaxy. We present a comparison of the time variation of the 
   abundance gradients in the Milky Way disk as determined from a sample
   of planetary nebulae, open clusters, cepheids and young objects, such 
   as stars in OB associations and HII regions. We conclude that the 
   [Fe/H] gradients as measured in open cluster stars strongly support 
   the time flattening of the abundance gradient as determined from O/H 
   and S/H measurements in planetary nebulae. This conclusion is also
   supported by the cepheid variables, for which very accurate gradients 
   and ages can be determined, and also by some recent estimates for OB 
   stars and HII regions. It is estimated that the average flattening rate 
   for the last 8 Gyr is in the range $0.005 - 0.010$ dex kpc$^{-1}$ Gyr$^{-1}$.

   }

   \authorrunning {Maciel et al.}
   \titlerunning {Time variation of radial abundance gradients}

   \maketitle
%

\section{Introduction}

It is now generally accepted that the radial abundance gradients observed 
in the Milky Way disk are among the main constraints of models of the 
chemical evolution of the Galaxy. The study of the gradients comprises the 
determination of their magnitudes along the disk, which include 
possible space variations, and their time evolution during the lifetime
of the Galaxy (see for example Henry \& Worthey \cite{henry}, Maciel 
\cite{maciel00} and Maciel \& Costa  \cite{mc03} for recent reviews).

The magnitudes of the gradients can be derived from a variety of 
objects, such as HII regions, early type stars, planetary nebulae, open
clusters etc. Recent investigations include Deharveng et al. 
(\cite{deharveng}), Andrievsky et al. (\cite{serguei5}), Friel et al. 
(\cite{friel02}) and Maciel et al. (\cite{mcu2003}), for HII regions, 
cepheid variables, open clusters and planetary nebulae, respectively. 
Average values are generally in the range $-$0.04 to $-$0.10 dex/kpc for 
the best observed element ratios, which are O/H and S/H in photoionized 
nebulae and Fe/H in stars.

The space variations of the gradients are more controversial. Some
flattening at large galactocentric distances is clearly discernible
in a sample of galactic planetary nebulae, as shown by Maciel \& Quireza
(\cite{mq99}) and more recently by Costa et al. (\cite{costa2004}),
on the basis of a study of nebulae located in the direction of the 
galactic anticentre. These results are supported by some work on HII 
regions (see for example V\'\i lchez \& Esteban \cite{vilchez}). 
The cepheid data are consistent with a flattened gradient near the 
solar neighbourhood, as suggested by Andrievsky et al. (\cite{serguei5}). 
On the other hand, no flattening has been observed in some recent 
studies of O, B stars in the galactic disk (see for example Smartt 
\cite{smartt}) or from planetary nebulae (Henry et al. \cite{henry04}),
although in the latter case generally flatter gradients have
been determined.

Probably the most interesting property of the gradients is their time
evolution, as it appears to be a very distinctive constraint of many 
recent chemical evolution models (cf. Tosi \cite{tosi}). As an example, 
models by Hou et al. (\cite{hou}) and Alib\'es et al. (\cite{alibes}) 
predict a continuous time flattening of the gradients, while models 
by Chiappini et al. (\cite{cmr2001}) are consistent with some 
steepening on a timescale of 3 to 10 Gyr. Therefore, it is extremely 
important to obtain observational constraints on the time evolution 
of the gradients, along with their magnitudes and possible space 
variations. Recently, Maciel et al. (\cite{mcu2003}, hereafter referred 
to as Paper~I) suggested that the O/H gradient has been flattening 
from roughly $-$0.11 dex/kpc to $-$0.06 dex/kpc during the last 9 Gyr, 
or from $-$0.08 dex/kpc to $-$0.06 dex/kpc in the last 5 Gyr. These 
results were obtained using a large sample of planetary nebulae for 
which accurate abundances are available, and for which the ages of 
the progenitor stars have been individually determined. As discussed 
by Maciel et al. (\cite{mcu2003}), the absolute ages derived are probably 
not accurate, but the relative ages of the stars are better determined, 
so that the time behaviour of the gradient can be derived, at least 
for the last 5 Gyr, which include most objects in the sample.

Along with planetary nebulae, open clusters are favorite objects to 
study the time evolution of abundance gradients, as they comprise a wide 
age bracket and have relatively well determined ages, based on detailed 
comparisons of theoretical isochrones and color magnitude diagrams
(see for example Friel \cite{friel95}, \cite{friel99}, and Phelps 
\cite{phelps} for recent reviews). The distances are also generally 
well determined, while the stellar metallicities, mostly derived by 
photometric techniques, are not as accurate as in the case of some 
elements in  photoionized nebulae, but nevertheless the gradients can 
be derived within a similar uncertainty, roughly 0.01 dex/kpc.

Cepheid variables are also very interesting objects to study 
abundance gradients, since their distances, ages and chemical
composition can be extremely well determined in comparison with
the other objects considered here. This can be observed in the recent
series of papers by Andrievsky and collaborators (Andrievsky
et al. 2002abc, \cite{serguei5}, and Luck et al. \cite{serguei4}).
Cepheid variables are young objects spanning a limited range in ages, 
and we will show in section~4 that this characteristic is crucial 
in the determination of the gradient at the present time.

In the present work, we take into account the results of Maciel 
et al. (\cite{mcu2003}, Paper I) and extend the discussion on the 
abundance gradients by (i) including S/H data as a tool in the study 
of the time variation of the gradients, (ii) adopting [Fe/H] $\times$ 
O/H and S/H conversions based on our previous work, and (iii) making 
a detailed estimate of the gradient from Cepheid data and of the 
Cepheid age distribution. Finally, (iv) we take into account some recent
determinations of the gradients from young objects, such as HII regions
and stars in OB associations, and show that a detailed comparison of
results based on samples of different objects, involving different
techniques and observational data, leads to a consistent interpretation
of the time variation of the radial abundance gradients in the 
galactic disk.

\section{Gradients from planetary nebulae}

\subsection{O/H gradients} 

An estimate of the time variation of the O/H radial gradient
in the galactic disk has recently been made by Maciel et al.
(\cite{mcu2003}). From the observed O/H abundances in a large
sample of nebulae, the [Fe/H] metallicity was determined on
the basis of a correlation with [O/H] abundances derived for 
disk stars. Here the brackets refer to abundances relative
to the Sun, as usual. An age-metallicity relation was used to 
estimate the ages of the progenitor stars, so that the temporal 
behaviour of the gradients could be derived. Two cases were 
considered (A and B), in which the sample was divided into three 
age groups, namely, Case~A: Group~I, with ages in the range 
$0 < t ({\rm Gyr}) < 3$, Group~II, for which 
$3 < t ({\rm Gyr}) < 6$, and Group~III, with $t > 6\ {\rm Gyr}$; 
and Case~B: Group~I, with ages in the range $0 < t ({\rm Gyr}) < 4$,
Group~II, for which $4 < t ({\rm Gyr}) < 5$, and Group~III, with 
$t  > 5\ {\rm Gyr}$. 

The results of case B, which are statistically more significant,
are shown in Table~\ref{gradpn}. The table gives the results for the 
O/H gradients  in the form $\log {\rm (O/H)} + 12 = a + b R$,
where $R$ is the galactocentric distance (kpc). The age groups
are defined in columns~1 and 2, and we have adopted 8~Gyr as
the upper age limit of the oldest group, since about 97\% of
the objects in our sample have ages lower than this limit. For each 
age group, the columns give the intercept $a$ with associated 
uncertainties (column~3), the gradient $b$ (dex/kpc) (column~4), the 
correlation coefficient $r$ and number of data points $n$ (columns~5 
and 6, respectively). These gradients, as well as other gradients derived 
in this paper, were calculated adopting a value $R_0 = 8.0$ kpc
for the solar galactocentric distance. Recent discussions on this
topic suggest $R_0$\ values in the range 7.6 -- 8.0 kpc (see for 
example Reid \cite{reid}, Maciel \cite{maciel93} and McNamara et al. 
\cite{mcnamara}). The adoption of different values within
this range would have a small effect on the gradients, so that 
our results can be safely compared with other
calculations using $R_0 = 7.6$ kpc (Maciel \& Quireza \cite{mq99})
or $R_0 = 7.9$ kpc (Andrievsky et al. \cite{serguei5} and Daflon
\& Cunha \cite{daflon}, see sections 4 and 5).

\begin{table*}
\caption[]{Abundance gradients from planetary nebulae, given as
$\log {\rm (X/H)} + 12 = a + b R$.}
\begin{flushleft}
\begin{tabular}{lllllll}
\hline\noalign{\smallskip}
Group  & Age (Gyr) & \ \ \ \ \ \ \  $a$  &\ \ \  $b$ (dex/kpc) &\ \ \  $r$ & $n$ & 
$d{\rm [Fe/H]}/ dR$  \\
\noalign{\smallskip}
\hline\noalign{\smallskip}
O/H gradients  & &                &                   &          &     &          \\
I    &  $0-4$  & $9.252\pm 0.064$ & $-0.047\pm 0.007$ &  $-0.64$ & 66  & $-0.056\pm 0.008$ \\
II   &  $4-5$  & $9.337\pm 0.027$ & $-0.089\pm 0.003$ &  $-0.94$ & 99  & $-0.107\pm 0.004$ \\
III  &  $5-8$  & $9.049\pm 0.072$ & $-0.094\pm 0.010$ &  $-0.75$ & 69  & $-0.113\pm 0.012$ \\
     &         &                  &                   &          &     &          \\
S/H gradients  & &                &                   &          &     &       \\
I    &  $0-4$  & $7.648\pm 0.132$ & $-0.080\pm 0.014$ &  $-0.66$ & 44  & $-0.096\pm 0.017$ \\
II   &  $4-8$  & $7.730\pm 0.091$ & $-0.113\pm 0.011$ &  $-0.76$ & 72  & $-0.136\pm 0.013$ \\
\noalign{\smallskip}
\hline
\end{tabular}
\end{flushleft}
\label{gradpn}
\end{table*}

The first three rows of Table~\ref{gradpn} show that the
O/H gradient flattens out in time, that is, the O/H gradient
becomes steeper along the sequence of groups I, II and III.
In this section, we will show that the observed flattening
of the gradients does not depend on the precise definition
of the age groups, which is largely arbitrary. To do this,
we have divided the PN sample into two age groups only, which 
we will call Group I, or \lq\lq younger\rq\rq, and Group II,
or \lq\lq older\rq\rq. Let $t_I$ be the upper age limit 
of Group~I, so that all PN progenitors having ages
$t \leq t_I$ will belong to Group~I, while those with
ages $t > t_I$ belong to Group~II. We have then adopted
several different values of $t_I$ in the range 
$3.0  < t_I {\rm (Gyr)} < 6.0$, and for each of these values
we have calculated  the O/H gradient of the corresponding
Groups I and II.

\begin{figure*}
   \centering
   \includegraphics[angle=-90.0,width=13.0cm]{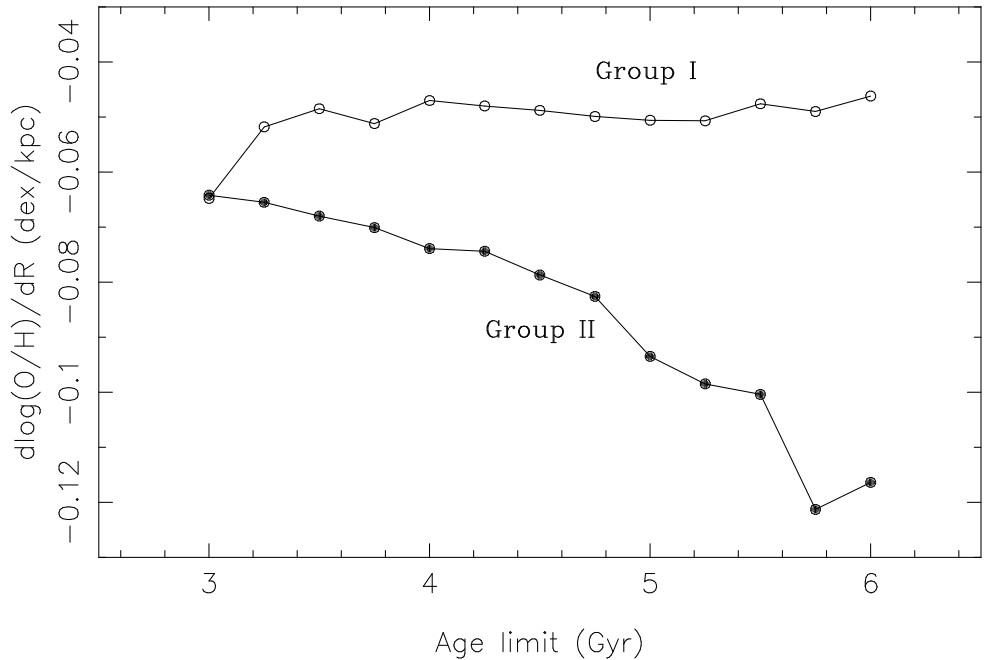}
      \caption{Time variation of the O/H gradient from planetary
      nebulae. The PN sample was divided into two
      age groups, Group I (\lq\lq younger\rq\rq), with ages lower
      than the age limit $t_I$, and Group II (\lq\lq older\rq\rq), 
      with ages higher than $t_I$. The plot shows the O/H gradient 
      (dex/kpc) of each group as a function of the upper age 
      limit of Group~I, $t_I$. The gradients of the younger Group~I 
      (open circles connected by lines) are always flatter than 
      those of the older Group~II (filled circles connected by lines).
      The best results apply for $t_I \simeq 4$ to 5 Gyr, approximately,
      for which the correlations are more meaningful.}
   \label{gradohv}
   \end{figure*}

Fig.~\ref{gradohv} shows the obtained O/H gradients as a function
of the age limit $t_I$. The gradients of Group~I are shown as
empty circles connected by lines, while those of Group~II 
are represented by filled circles. From this figure, it 
becomes immediately clear that the younger Group~I 
has systematically flatter gradients than the older Group~II, 
irrespective of the adopted age limit, that is,
the O/H gradient appears to be flattening out in time, no
matter how one defines Groups~I and~II. It can
also be seen that the difference between the gradients of
Groups I and II increases with $t_I$, which reflects the 
fact that for larger $t_I$ values only the oldest objects,
which show the steepest gradients, are allocated to Group~II.

Naturally, for values of $t_I$ close to the minimum age limit,
$t_I \simeq 3$ in Fig.~\ref{gradohv}, most objects are allocated 
to Group~II, so that Group~I becomes too small. Conversely, for
$t_I \simeq 6$ most objects are in Group~I, and Group~II becomes 
underpopulated. Therefore, in these cases the correlation 
is less meaningful, so that the best results are obtained 
for $t_I \simeq 4$ to 5 Gyr, approximately.

From the results in Fig.~\ref{gradohv} and Table~\ref{gradpn}, 
it can be seen that the O/H gradient flattens out from the
older groups to the younger ones. Although the choice of the age 
groups is arbitrary, in all cases considered by us the flattening 
of the gradient is apparent. As discussed by Maciel et al. 
(\cite{mcu2003}), even though the {\it absolute} ages may be in error, 
it is unlikely that the {\it relative} ages are incorrect, so that 
the flattening of the gradients is probably real. Also, by
considering two age groups only, the probability of placing
a given object in the wrong group is relatively small. Furthermore, 
Maciel et al. (\cite{mcu2003}) performed an independent calculation 
of the ages, on the basis of a relationship between the N/O abundance
and the progenitor star mass, which led to the original stellar
mass on the main sequence and therefore to another estimate of
the age, with results similar to those shown in Table \ref{gradpn}.

In order to compare the PN derived gradients with results from other
objects, especially open cluster stars and cepheids, it is interesting
to convert the observed O/H gradient into [Fe/H] gradients. In fact, 
the O/H and [Fe/H] are expected to be similar, but not identical, 
which can be concluded by an inspection of the relation between 
the [O/Fe] ratio and the metallicity [Fe/H] in the galactic disk
(see for example the discussion by Maciel \cite{maciel02}).
[Fe/H] gradients cannot be determined directly from planetary
nebulae, as the iron lines are weak and a sizable fraction
of this element is probably locked up in grains. However, 
a relation between the iron and oxygen abundances can be 
derived from observed properties of the stellar populations 
in the galactic disk. A detailed discussion on the metallicities 
and radial gradients from a variety of sources such as HII 
regions, hot stars and planetary nebulae in the galactic disk 
has been recently presented by Maciel (\cite{maciel02}).
According to this analysis, an independent [O/Fe] $\times$
[Fe/H] relation has been obtained for the galactic disk,
from which we can write approximately

   \begin{equation}
     {\rm [Fe/H]} = \gamma + \delta \ (\log {\rm O/H} + 12) \ ,
     \label{fehoh}
   \end{equation}

\noindent
where $\gamma$ and $\delta$ are coefficients essentially 
independent of the galactocentric distance. 
An average value $\delta \simeq 1.2$ was determined by
Maciel (\cite{maciel02}), which is slightly lower than the
value adopted for the solar neighbourhood by Maciel et al. 
(\cite{mcu2003}), $\delta \simeq 1.4$. In fact, this parameter
may show some time dependence, as we will see from a comparison
of the PN gradients with those derived from open clusters,
but it is typically in the range  $\delta \simeq 1.0 - 1.5$,
so that the main conclusions of this paper are not affected
by this variation. Since both O/H and [Fe/H] gradients are 
assumed to be linear, it is easy to see that

   \begin{equation}
      {d{\rm [Fe/H]}\over dR} \simeq \delta \, 
   {d\log {\rm (O/H)} \over dR} \ ,
     \label{relgrad}
   \end{equation}

\noindent
which can be applied to the O/H gradients of Table~\ref{gradpn}.
The derived [Fe/H] gradients are given in the last column
of Table~\ref{gradpn}. 

It should be mentioned that we have taken into account the new 
determinations of the solar oxygen abundance, which has been 
revised downwards following new 3D model atmospheres 
(Allende-Prieto et al. \cite{allende}, Asplund \cite{asplund1} 
and Asplund et al. \cite{asplund2}). We have adopted here 
$\log {\rm (O/H)_\odot} + 12 = 8.7$.

\subsection{S/H gradients} 

Radial gradients involving the S/H ratio can also be derived
from planetary nebulae, as shown for example in our earlier
work (Maciel \& K\"oppen \cite{mk94} and Maciel \& Quireza
\cite{mq99}), as well from HII regions (see for example
Afflerbach et al. \cite{afflerbach}). The obtained values are
similar and probably slightly steeper than the O/H gradient.

We have considered the objects in the sample of Maciel et al. 
(\cite{mcu2003}) for which reliable sulfur data are available
and applied the same procedure as for the O/H ratio. Lists
of objects and corresponding abundances are given in Maciel
\& Quireza (\cite{mq99}) and Maciel et al. (\cite{mcu2003}),
which were supplemented by the new results for anticentre
nebulae by Costa et al. (\cite{costa2004}). In view of the
weakness of the sulfur lines in photoionized nebulae, and
the possible effect of the S$^{+++}$ ion (see Henry et al.
\cite{henry04} for a discussion), the derived abundances 
are somewhat less accurate than for oxygen, which is reflected 
by a smaller sample and a relatively larger dispersion. 
For this reason, we decided to divide the PN sample into two 
age groups only, namely Group~I and II, which we will refer 
to as the \lq\lq younger\rq\rq\ and \lq\lq older\rq\rq\ groups,
respectively. This is basically the same procedure adopted
in the previous sub-section, which led to Fig.~\ref{gradohv}.
Since all objects in our sample have ages lower than 7 Gyr, 
we have adopted several values for the upper age limit of the 
younger group in the range $3.0  < t_I {\rm (Gyr)} < 6.0$.

   \begin{figure*}
   \centering
   \includegraphics[angle=-90.0,width=13.0cm]{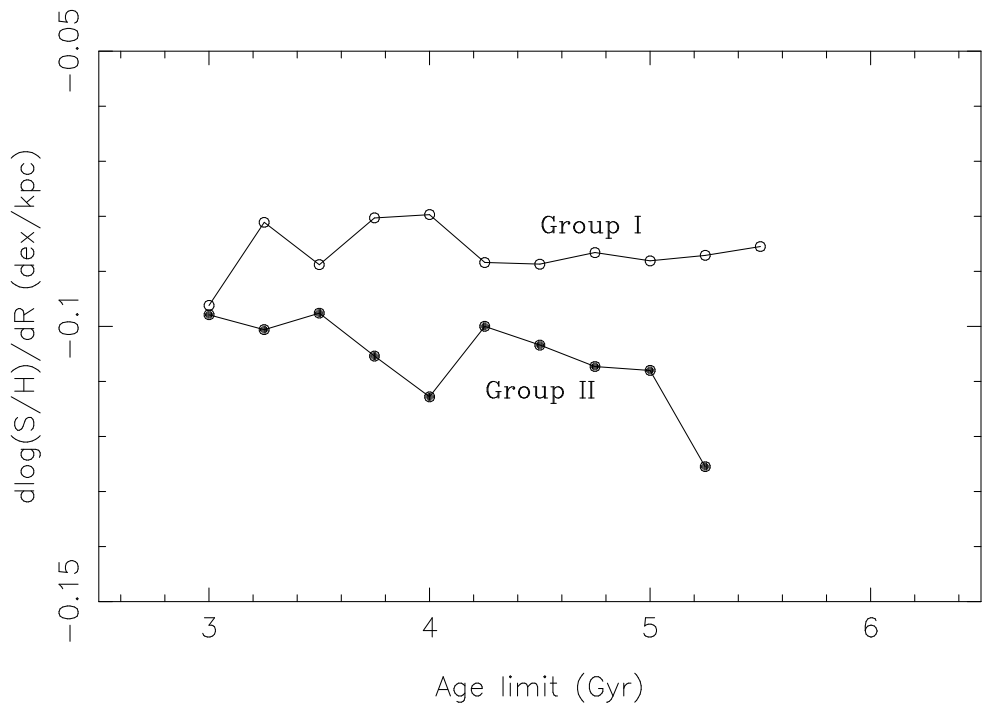}
      \caption{The S/H gradient (dex/kpc) from planetary nebulae,
      divided into two age groups, as a function of the upper age 
      limit of Group~I. The gradients of the younger Group~I (open 
      circles connected by lines) are always flatter than those 
      of the older Group II (filled circles connected by lines).}
   \label{gradshv}
   \end{figure*}

   \begin{figure*}
   \centering
   \includegraphics[angle=-90.0,width=13.0cm]{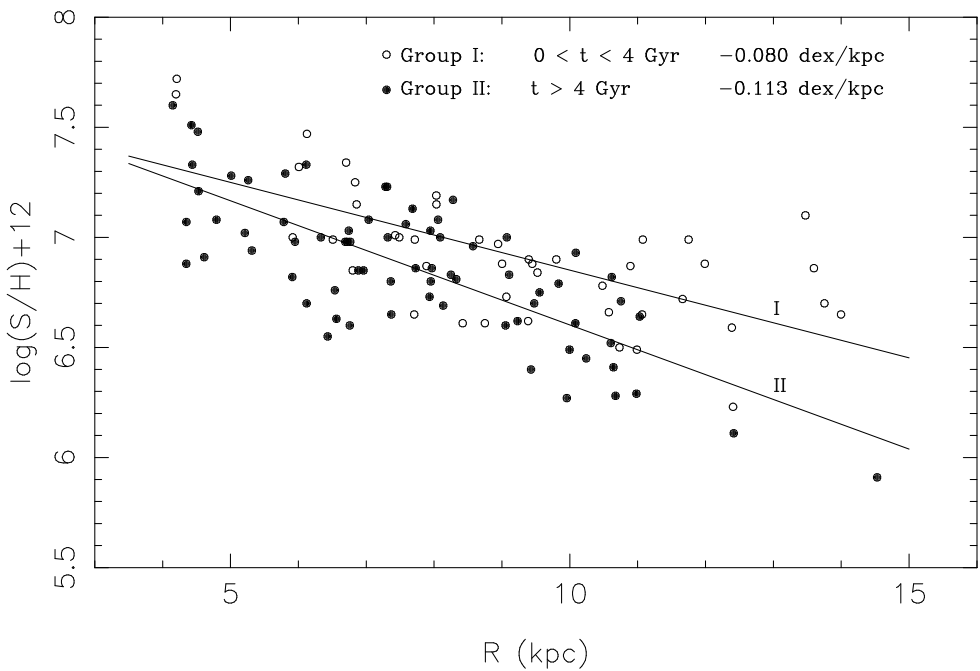}
      \caption{The S/H abundance gradient from planetary nebulae,
      divided into two age groups: Group I ($t < 4$ Gyr,  empty circles) 
      and Group II ($t > 4$ Gyr, filled circles). The corresponding 
      gradients (dex/kpc) are shown in the figure. The adopted value 
      of the galactocentric distance of the Sun is $R_0 = 8.0$ kpc.}
   \label{gradsh}
   \end{figure*}

Fig.~\ref{gradshv} shows the derived S/H gradients as a function of
the upper age limit for Group~I, and can be compared with Fig.~\ref{gradohv}. 
It can be seen that the younger group I (empty circles) shows gradients 
systematically flatter than the older Group~II (filled circles), in 
agreement with the results for oxygen. Again, for $t_I \leq 3$ Gyr and 
$t_I \simeq 6$ Gyr the samples become too small, and no meaningful results 
can be obtained, so that the best results occur for ages in the
range $3.5 < t_I ({\rm Gyr}) < 5$.  As an example, Fig.~\ref{gradsh} 
shows the derived gradients for the case $t_I = 4.0$ Gyr, 
that is, in this case the objects in the younger Group I have ages lower 
than 4 Gyr, while the older group II has ages larger than 4 Gyr.
The corresponding values are given in the last two rows
of Table~\ref{gradpn} for the two age groups. Again the 
flattening of the gradients with time is apparent.

In order to convert the S/H gradients into [Fe/H] gradients, we
have followed the same procedure as for oxygen, and have first
obtained a correlation between the S/H and O/H abundance ratios.
Such a correlation is predicted by nucleosynthetic models of
intermediate mass stars, since S and O are not basically altered
by the evolution of the progenitor stars of planetary nebulae.
In fact, several people have independently obtained relatively
tight correlations between S/H and O/H, as well as for Ne/H and
Ar/H (see for example K\"oppen et al. \cite{kas}, Maciel \& 
K\"oppen \cite{mk94}, Costa et al. \cite{costa2004}, Henry et al.
\cite{henry04}). From these references, and especially from
the results of our own group (Maciel \& K\"oppen \cite{mk94}, 
Maciel \& Quireza \cite{mq99}, Costa et al. \cite{costa2004}),
we conclude that the slope is very close to unity, so that we can 
apply for the S/H gradient the same correction factor as for O/H, 
namely, $d{\rm [Fe/H]}/dR \simeq 1.2\ d\log{\rm (S/H)}/dR$.
The corresponding [Fe/H] gradients are also given in the last column
of Table~\ref{gradpn}.

\section {Gradients from open clusters}

As mentioned in the Introduction, open clusters are favorite
objects for which the determination of abundance
gradients is possible (Janes \cite{janes}, Friel \cite{friel95}, 
and Phelps \cite{phelps}). In a recent work, Friel and 
collaborators presented a sample of 39 open clusters for 
which metallicities, distances and ages have been determined in 
a homogeneous and consistent way (Friel et al. \cite{friel02}, 
see also Friel \cite{friel05}). On the basis of a updated
abundance calibration of spectroscopic indices in a 
sample containing 459 stars, Friel et al. (\cite{friel02})
derived an average [Fe/H] gradient of $-0.06$ dex/kpc
in a range of galactocentric distances of 7 to 16 kpc.
Taking into account age groups defined in several different
ways, they concluded that there is an indication of some
time flattening of the gradients. For example, for clusters 
with ages under 2 Gyr, the derived gradient is
$-0.023$ dex/kpc; for ages between 2 Gyr and 4 Gyr, the gradient
is $-0.053$ dex/kpc, and for those older than 4 Gyr they 
obtained $-0.075$ dex/kpc.

More recently, new catalogues and compilations of open clusters 
have become available (Chen et al. \cite{chen}, Dias et al. 
\cite{wilton}), which include space and kinematical data, as well
as metallicities and estimates of the ages. Chen et al. 
(\cite{chen}) assembled a total of 119 clusters, for which distances
and metallicities are available, which led to an
average gradient of $d$[Fe/H]$/dR \simeq -0.063\,$dex/kpc
for the whole sample, similar to the value derived for
the homogeneous sample by Friel et al. (\cite{friel02}).
Taking into account two different age groups (ages $< 0.8$ Gyr 
and $\geq 0.8$ Gyr, respectively), Chen et al. (\cite{chen}) 
concluded that the iron gradient was steeper in the past, 
supporting the earlier conclusion by Friel et al. (\cite{friel02}).

In order to obtain a more accurate comparison between
the [Fe/H] gradients  from open clusters and the planetary
nebula data, we have used both the homogenous sample by
Friel et al. (\cite{friel02}) and the compilation by
Chen et al. (\cite{chen}) and rederived the [Fe/H]
gradients, taking into account several possibilities of
defining the age groups. Following the same procedure as
for the O/H and S/H gradients in planetary nebulae,
we have divided the open cluster stars into two age groups,
varying the upper age limit of the younger Group~I.

   \begin{figure*}
   \centering
   \includegraphics[angle=-90.0,width=13.0cm]{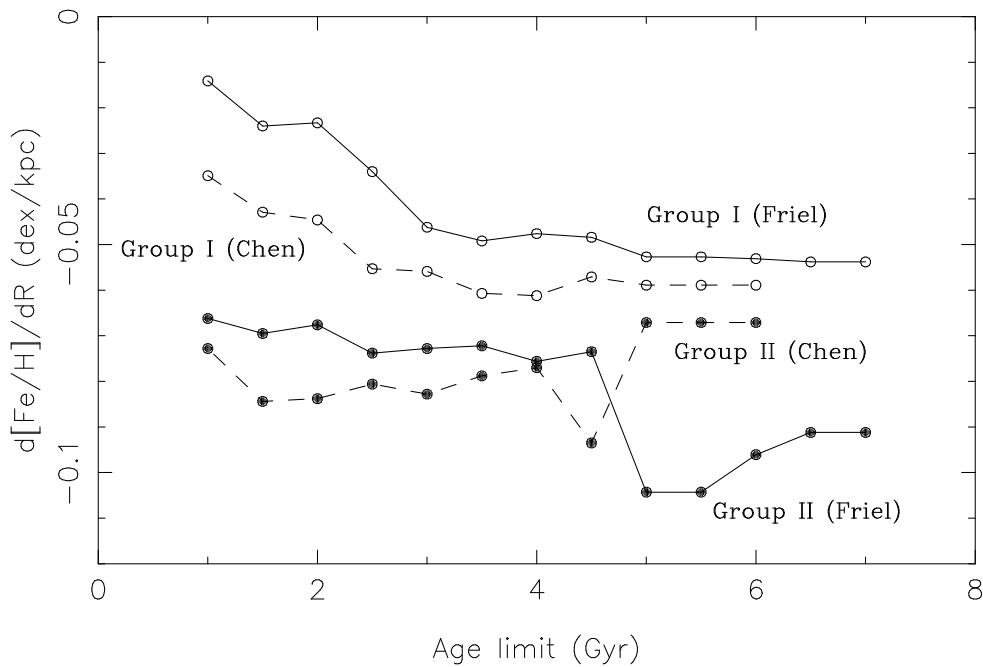}
      \caption{The [Fe/H] gradient (dex/kpc) from the samples of
       open clusters by Friel et al. (\cite{friel02}, solid lines),
      and Chen et al. (\cite{chen}, dashed lines), divided into 
      two age groups, as a function of the upper age 
      limit of Group~I. As for the case of planetary nebulae, 
      the gradients of the younger Group~I (open circles) are always 
      flatter than those of the older Group~II (filled circles).}
   \label{frielchen}
   \end{figure*}

Fig.~\ref{frielchen} shows the [Fe/H] gradient as a
function of the upper age limit of Group~I for the
samples of Friel et al. (\cite{friel02}, solid lines)
and Chen et al. (\cite{chen}, dashed lines). Again
we observe the same behaviour as in Fig.~\ref{gradohv}
and Fig.~\ref{gradshv}, in the sense that the gradients
of Group~I objects are always flatter than for Group~II.
We can obtain a more detailed result dividing the
open clusters into three age groups, as done by Chen et
al. (\cite{chen}). A representative example is given in
Table~\ref{gradoc}, where the groups are defined in
columns~1 and 2, and the parameters of the fits are
shown in columns~3 to 6. The upper age limit is again
8~Gyr, which includes about 96\% of the total sample
of open clusters, in good agreement with the recent 
estimates for the ages of the oldest open clusters by
Salaris et al. (\cite{salaris}). We have restricted 
our analysis to galactocentric distances $R < 16\,$kpc, 
and excluded the object Berkeley 29, which does not satisfy 
this criterium (see Chen et al. \cite{chen} for details). 
The total number of clusters in the samples are 39 
and 118 for the objects of Friel et al. (\cite{friel02})
and Chen et al. (\cite{chen}), respectively. It should
be noted that both samples are not completely independent,
as most of the objects in the Friel sample are also contained in
the compilation by Chen et al. (\cite{chen}), although the
adopted parameters (ages and metallicities) are often
different. However, the fact that the homogeneous sample 
by Friel et al. and the compilation by Chen et al. lead 
essentially to the same results shows that the lack of 
homogeneity in the determination of key parameters such 
as the ages and metallicities in the larger sample by Chen 
et al. does not introduce any significant uncertainty 
relative to the homogeneously obtained sample by Friel 
et al. (\cite{friel02}).

\begin{table*}
\caption[]{Abundance gradients from open clusters, given as
[Fe/H] $ = a + b R$.}
\begin{flushleft}
\begin{tabular}{llllll}
\hline\noalign{\smallskip}
Group  & Age (Gyr) & \ \ \ \ \ \ \  $a$  &\ \ \  $b$ (dex/kpc) &\ \ \  $r$ & $n$ \\
\noalign{\smallskip}
\hline\noalign{\smallskip}
Friel &         &                   &                   &          &     \\
I    &  $0-2$   & $-0.023\pm 0.169$ & $-0.023\pm 0.019$ &  $-0.33$ & 15  \\
II   &  $2-5$   & $0.279\pm 0.156$  & $-0.058\pm 0.014$ &  $-0.77$ & 14  \\
III  &  $5-8$   & $0.715\pm 0.327$  & $-0.104\pm 0.035$ &  $-0.73$ & 10  \\
     &          &                   &                   &          &     \\
Chen &          &                   &                   &          &     \\
I    &  $0-0.8$ & $0.185\pm 0.103$  & $-0.024\pm 0.012$ &  $-0.22$ & 80  \\
II   &  $0.8-2$ & $0.467\pm 0.175$  & $-0.067\pm 0.018$ &  $-0.69$ & 18  \\
III  &  $2-8$   & $0.681\pm 0.218$  & $-0.084\pm 0.020$ &  $-0.71$ & 20  \\
\noalign{\smallskip}
\hline
\end{tabular}
\end{flushleft}
\label{gradoc}
\end{table*}

Taking into account the larger sample by Chen et al. 
(\cite{chen}), we notice that most of the age groups have 
galactocentric distances roughly in the range $R \simeq 6$ to 
15 kpc, which is wide enough so that the derived gradients 
are representative of the galactic disk. However, there is 
some tendency for the youngest clusters to be located at 
galactocentric distances closer than about 10 kpc, which 
probably reflects the fact that older clusters are destroyed 
by collisions with molecular clouds in the inner Galaxy. 
Therefore, the distribution of the young clusters presents a 
more limited range of galactocentric distances, so that the 
derived gradients may be artificially flat, which is also 
reflected by the lower correlation coefficients of Group~I 
in both samples, as can be seen in Table~\ref{gradoc}. In order 
to investigate the amount by which this affects the results for the 
gradients at the present time, we will consider in the next section 
the abundance gradients as derived from cepheid variable stars.

\section {Cepheid variables}

\subsection{The [Fe/H] gradient from cepheids}

Cepheid variables have a distinct role in the determination of
radial abundance gradients and their time variations for
a number of reasons. First, they are usually bright enough  
that they can be observed at large distances, providing 
accurate abundances; second, their distances are generally
well determined, as these objects are often used as distance
calibrators; third, their ages are also well determined, on
the basis of relations involving their periods, luminosities,
masses and ages. Of course, they generally have ages close to
a few hundred million years, so that they cannot be used alone 
in the study of the time variations of the gradients, but define 
instead an accurate benchmark for the value of the gradient at 
their age bracket.

Recently, accurate abundances of several elements have been derived
by Andrievsky et al. (2002abc, 2004) and Luck et al. (\cite{serguei4}) 
for a large sample of galactic cepheids. The total sample includes 
over 120 objects, and abundances of about 20 heavy elements have 
been derived, including C, O and Fe up to Nd and Eu. A clear negative 
gradient is evident from their results (see for example fig. 3 of 
Andrievsky et al. 2004), which is roughly similar to the gradients 
derived from planetary nebulae and HII regions. Regarding the [Fe/H] 
abundances, which are better determined, they suggest a three zone 
gradient, amounting to
$d$[Fe/H]/$dR \simeq -0.128$ dex/kpc for $R = 4.0 - 6.6$ kpc
(zone I);
$d$[Fe/H]/$dR \simeq -0.044$ dex/kpc for $R = 6.6 - 10.6$ kpc
(zone II); and
$d$[Fe/H]/$dR \simeq 0.004$ dex/kpc for $R = 10.6 - 14.6$ kpc
(zone III), where they have taken $R_0 = 7.9$ kpc. A change of
slope near the solar galactocentric radius has also been
suggested by Caputo et al. (\cite{caputo}) for cepheids and
Twarog et al. (\cite{twarog}) for open clusters. Also,
the flattening of the gradient at large galactocentric
distances is probably real, and has been observed by our
group from planetary nebulae (cf. Costa et al. \cite{costa2004}
and Maciel \& Quireza \cite{mq99}). The division
in three zones as adopted by Andrievsky et al. (\cite{serguei5})
is based on a smoothing of the data, which makes no assumptions 
about the structure of the gradient. However, the three zones 
are not equally populated, zones~I and III being undersampled 
relatively to zone~I, which may affect the gradients. On the
other hand, an average gradient valid for all zones can represent 
very well the cepheid data, as we will show, and is more 
adequate in order to make comparisons with other data.

\begin{figure*}
   \centering
   \includegraphics[angle=-90,width=13.0cm]{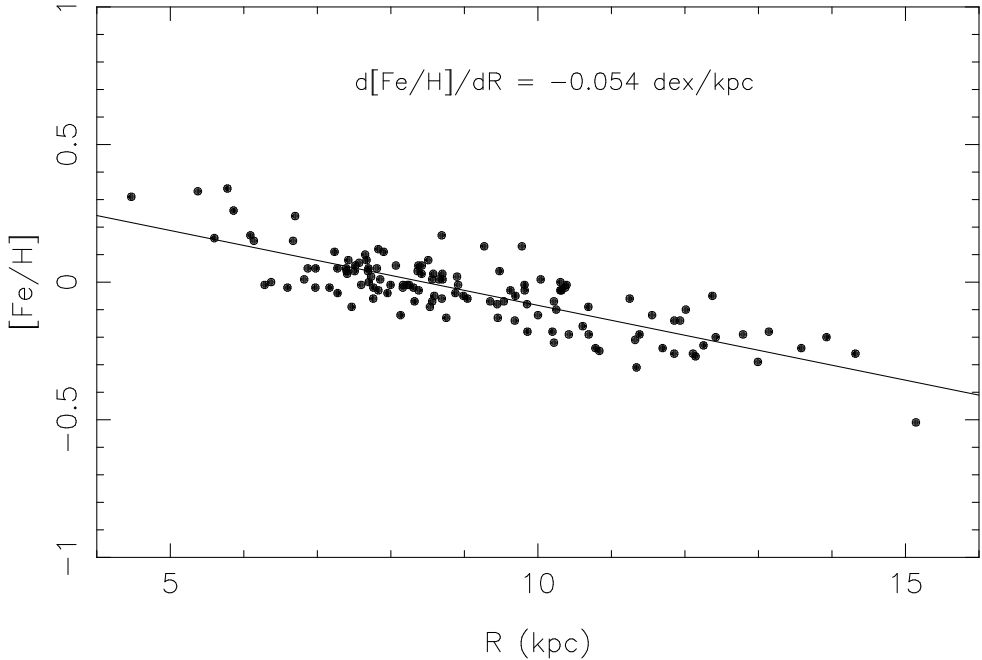}
      \caption{The [Fe/H] abundance gradient from cepheid variable
      stars. The correlation coefficient is $r = -0.82$, and a total
      of 127 stars have been included. The adopted value for the 
      galactocentric distance of the Sun is $R_0 = 8.0$ kpc.}
   \label{gradceph}
   \end{figure*}

We have taken into account the complete sample of
Andrievsky et al. (2002abc, 2004) and Luck et al.
(\cite{serguei4}) and rederived the [Fe/H] gradient
adopting $R_0 = 8.0$ kpc. We have obtained the following
result:

   \begin{equation}
      {\rm [Fe/H]} = 0.459 \pm 0.032 - (0.054 \pm 0.003) \ R \, 
     \label{eqceph}
   \end{equation}

\noindent
with a correlation coefficient $ r = -0.82$ and a total of
$n = 127$\ stars. These results are shown in Fig.~\ref{gradceph},
again for $R_0 = 8.0$ kpc, with no significant changes
for $7.6 < R_0 {\rm (kpc)} < 8.0$. The derived gradient 
is also similar to the average value one would get combining
the three zones considered by Andrievsky et al. (\cite{serguei5}).
It should be noted that the correlation shown by Fig.~\ref{gradceph}
is very well defined, as can be seen by the large correlation
coefficient obtained, which also reflects the high accuracy
of the spectroscopic cepheid data.

\subsection{Age distribution of cepheids}

In order to compare the average gradients derived in the
previous subsection with those from planetary nebulae
and open cluster stars, it is necessary to estimate the
age distribution of our sample of cepheid variables.
We have accomplished that using two different methods,
which we shall call Method 1 and Method 2. In both cases
we have used the periods, effective temperatures and
gravities given by Andrievsky et al. (2002abc, 2004) and
Luck et al. (\cite{serguei4}). The main difference of
the methods lies essentially in the determination of the 
stellar luminosities.

\bigskip\noindent
{\bf Method 1} - We have first derived a very simple 
period-luminosity relation for galactic cepheids using data
from Cox (\cite{cox}), which can be written as 

   \begin{equation}
     \log {L_\ast \over L_\odot} = 2.291 + 1.266 \log P \ ,
     \label{cox}
   \end{equation}

\noindent
where the period $P$ is in days. Since 
$L_\ast = 4 \pi R_\ast^2 \sigma T_{eff}^4$ and 
$g = G M_\ast/R_\ast^2$, where $R_\ast$ and $M_\ast$ are
the stellar radius and mass, respectively, and $T_{eff}$
is the effective temperature, the stellar mass can be 
derived for each object. The age follows from the calibration 
by Bahcall \& Piran (\cite{bahcall}) given by

   \begin{equation}
     \log t = 10 - 3.6 \log {M_\ast \over M_\odot} +
     \biggl(\log {M_\ast \over M_\odot}\biggr)^2  \ ,
     \label{bp83}
   \end{equation}

\noindent
where the age $t$ is in yr. An alternative mass--age
relationship can be obtained from Binney \& Merrifield
(\cite{binney}, p. 280, eqs. 5.5 and 5.6), from which we have

   \begin{equation}
     \log t = 9.75 - 2.5 \log {M_\ast \over M_\odot}  \ ,
     \label{bm98}
   \end{equation}
\noindent
which is valid for masses in the range
$2 < M_\ast/M_\odot < 20$. Here again the age $t$ is in yr.

\bigskip\noindent
{\bf Method 2} - The second method uses the metallicity--dependent 
period--luminosity relation recently derived by 
Groenewegen et al. (\cite{martin}), which can be
written for galactic cepheids as

   \begin{equation}
     \log {L_\ast \over L_\odot} = 2.432 + 1.092 \log P + 
     0.254 \ {\rm[Fe/H]} - 0.4 \ BC.
     \label{groenewegen}
   \end{equation}
\noindent
 The bolometric correction $BC$ was approximated as a function
of the effective temperature from Cox (\cite{cox}) as

   \begin{equation}
     BC = a_1 + a_2 \log T_{eff} + a_3 (\log T_{eff})^2  \ ,
     \label{bolometric}
   \end{equation}
\noindent
where $a_1 = -899.14547$, $a_2 = 477.87397$, $a_3 = -63.48878$.
Once the stellar luminosity has been obtained, the same procedure
as in method 1 is observed in order to derive the stellar ages.

\begin{figure*}
   \centering
   \includegraphics[angle=-90,width=13.0cm]{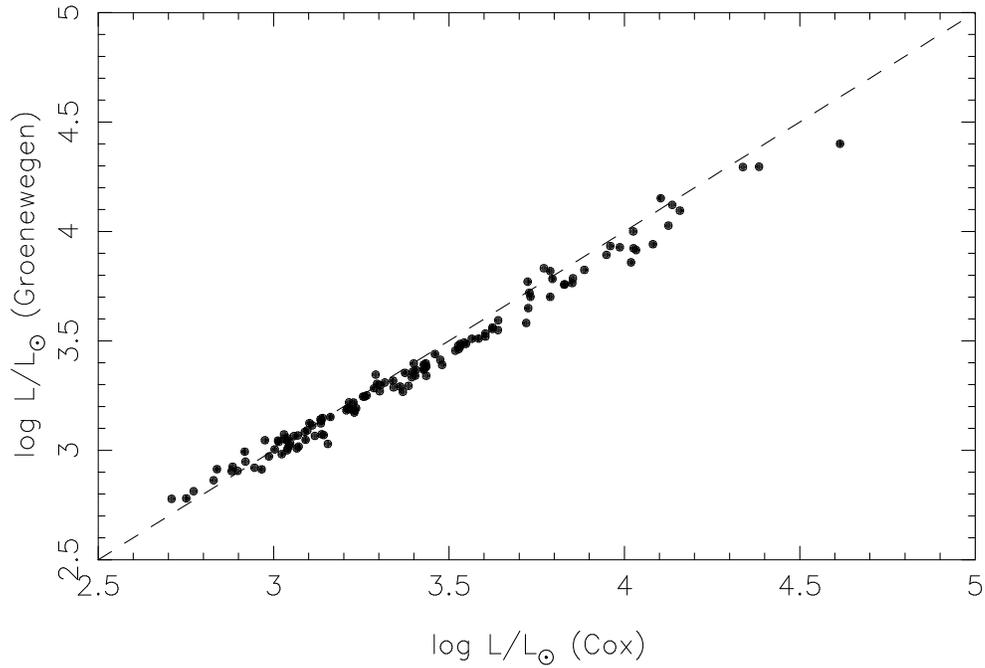}
      \caption{A comparison of the luminosities calculated using
      Method 1 (eq. \ref{cox}) from Cox (\cite{cox}) and
      Method 2 (eq. \ref{groenewegen}) from Groenewegen et
      al. (\cite{martin}).}
   \label{lumlum}
   \end{figure*}

A comparison of the luminosities derived from methods 1 and 2 
for the stars in our sample is shown in Fig.~\ref{lumlum}.  
It can ben seen that the agreement is very good, within 11\%
in average, so that the correction due to the metallicity effects
is small for all stars in our sample.

\begin{figure*}
   \centering
   \includegraphics[angle=-90,width=13.0cm]{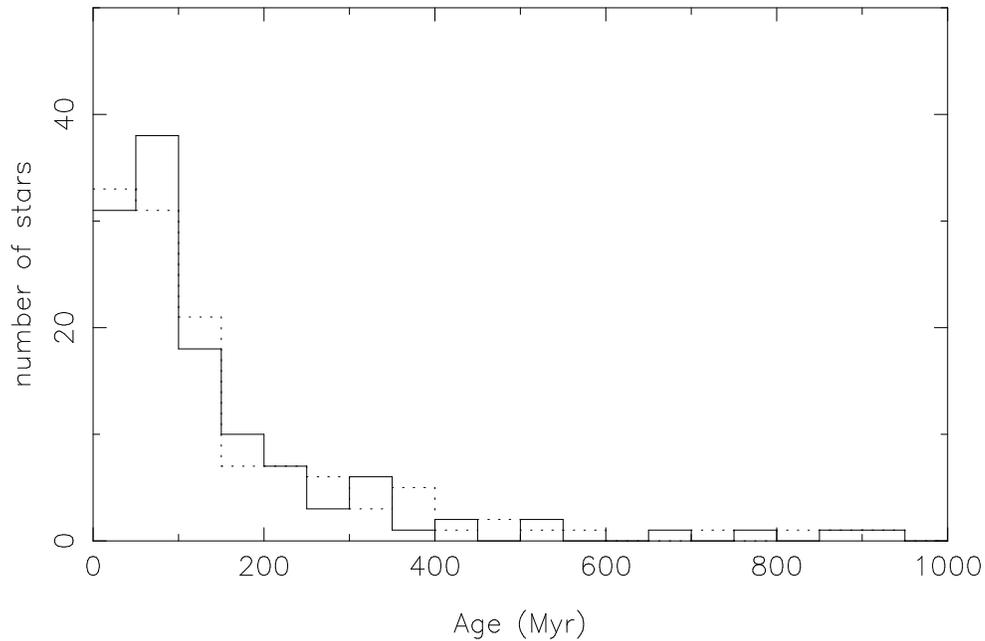}
      \caption{A comparison of the age distributions of cepheid
      variables using ages from Bahcall \& Piran (\cite{bahcall})
      [solid line, see eq. \ref{bp83}] and Binney \& Merrifield 
      (\cite{binney}) [dotted line, see eq. \ref{bm98}]. The
      period-luminosity relation is from Groenewegen et al.
      (\cite{martin}), [see eq. \ref{groenewegen}].}
   \label{histceph}
   \end{figure*}

The ages derived from methods 1 and 2 are then very similar,
the average difference between them being less than 20\% in
all cases.  This can be seen in the histograms shown in 
fig. \ref{histceph}, where ages using Bahcall \& Piran 
(\cite{bahcall}, eq. \ref{bp83}, solid line) or 
Binney \& Merrifield (\cite{binney}, eq. \ref{bm98}, dotted
line) have been used. The adopted period--luminosity relation
is eq. \ref{groenewegen}, from Groenwegen et al. (\cite{martin}).

The maximum ages calculated from both methods for the stars 
in our sample is about 1 Gyr, but most objects are much younger, 
as can be seen from fig. \ref{histceph}. In fact, about 80\%
of the stars have ages under 200 Myr, and 96\% of the cepheids
are younger than 700 Myr, so that the average gradient of the 
whole sample can be considered as representative of the 
\lq\lq young objects", as compared with most planetary nebulae
and open cluster stars. In the following, we will then adopt
a gradient of $d$[Fe/H]$dR = - 0.054 \pm 0.003 $ dex/kpc
for cepheid variables in  the approximate age range of
0 to 700 Myr.

\section{Young objects}

In order to obtain a comprehensive analysis of the time variation
of the abundance gradients, it is interesting to take into account 
some  very young objects, such as HII regions and OB stars in associations 
and young clusters, which better represent the youngest population 
for which meaningful gradients can be derived. 

\subsection{HII regions}

HII regions are classic objects in the determination of abundance
gradients, both in the Milky Way and in other galaxies (see for
example Henry \& Worthey \cite{henry} for a recent review). 
Several people have derived O/H gradients from
galactic HII regions (see for example Vilchez \& Esteban \cite{vilchez},
Afflerbach et al. \cite{afflerbach}, Deharveng et al. \cite{deharveng}
and Pilyugin et al. \cite{pilyugin} for recent determinations).
Generally, the most reliable determinations are those based on
measured [OIII] $\lambda$4959,5007/4363 lines, so that the
electronic temperature can be derived. The average values of the
O/H gradients are in the range $-0.040$ to $-0.070$ dex/kpc,
the flatter end being preferred by the most recent papers of
Deharveng et al. (\cite{deharveng}) and Pilyugin et al. (\cite{pilyugin}).
This derived range is considerably larger than the formal
uncertainties of each determination, which is typically of the 
order of 0.005 dex/kpc. Therefore, we will adopt here a conservative
average value given by $d\log {\rm (O/H)}/dR \simeq -0.055 \pm 0.015$ 
dex/kpc for the oxygen gradient from HII regions. In order to
estimate the corresponding [Fe/H] gradient, we will adopt
the same correcion factor as for the planetary nebulae,
so that we have $d{\rm [Fe/H]}/dR \simeq -0.066 \pm 0.018$ dex/kpc.
Since typical HII regions have ages in the Myr range, we may safely
assume that they are representative of the present day gradients.

\subsection{OB stars}

A sample containing 69 stars in OB associations, open clusters and
HII regions has been recently analysed by Daflon \& Cunha (\cite{daflon}),
on the basis of self-consistent non-LTE models with a homogeneous
set of stellar parameters. Abundances of the elements C, N, O, Mg, Al,
Si and S have been obtained, for which radial abundance gradients have
been computed. The sample objects have generally well determined 
distances, and they are all young, with ages under 50 Myr, so that
the derived gradients can also be considered as representative of 
the present day abundances in the galactic disk. The average gradient
derived by Daflon \& Cunha (\cite{daflon}) for all elements is
$-0.042\pm 0.007$ dex/kpc adopting $R_0 = 7.9$ kpc, which does not
change for $R_0 = 8.0$ kpc. For oxygen, they obtain a gradient of
$-0.031\pm 0.012$ dex/kpc, so that applying the same correction factor 
of 1.2 as before, we obtain a gradient of $d{\rm [Fe/H]} = 
-0.037\pm 0.014$\ dex/kpc, which is very close to the average value 
derived by Daflon \& Cunha (\cite{daflon}).

\section {Discussion}

\begin{figure*}
   \centering
   \includegraphics[angle=-90.0,width=13.0cm]{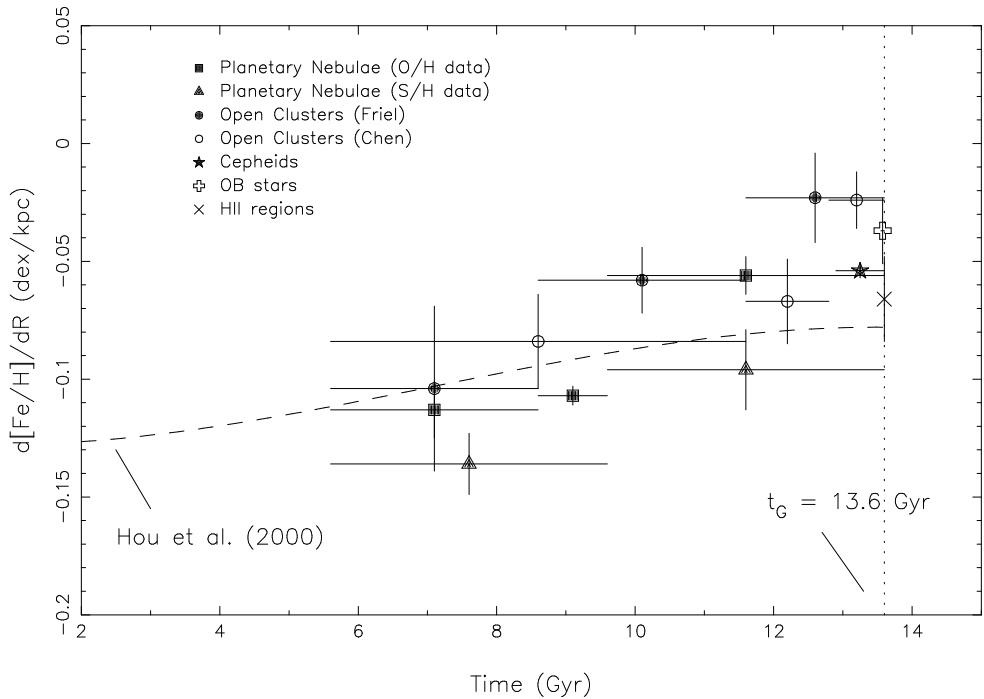}
      \caption{Time variation of the [Fe/H] abundance gradient (dex/kpc). 
      The converted [Fe/H] gradients from planetary nebulae are shown as 
      filled squares (O/H data) and filled triangles (S/H data). The open 
      cluster results are shown as filled circles (data from Friel et al. 
      \cite{friel02}) and empty circles  (data from Chen et al. \cite{chen}). 
      The [Fe/H] gradient from cepheid variables is shown as a filled star. 
      Also included are the following young objects: OB stars in associations 
      etc. form Daflon \& Cunha (\cite{daflon}, thick cross) and HII regions 
      (x sign). The dotted vertical line shows the adopted age of the 
      galactic disk, $t_G = 13.6$ Gyr, and the dashed line represents the 
      time evolution of the [Fe/H] gradient according to the theoretical 
      models of Hou et al. (\cite{hou}).}
   \label{vargrad}
   \end{figure*}

An inspection of the O/H and S/H gradients from planetary nebulae shown in 
Table \ref{gradpn} and the [Fe/H] gradients from open clusters (Table 
\ref{gradoc}) confirms that the average gradients have been flattening 
out for the last approximately 8 Gyr. These results are better seen in 
Fig.~\ref{vargrad}, where we plot the [Fe/H] gradients (dex/kpc) as a 
function of time, adopting $t_G = 13.6$\ Gyr for the age of the galactic 
disk. The gradients from planetary nebulae, obtained after converting
the O/H gradients as discussed in section~2 are shown as filled squares, 
while those derived from S/H data are shown as filled triangles. The 
horizontal error bars reflect the total age span as given in Table~\ref{gradpn},
and the vertical error bars are the calculated uncertainties of the
least squares fits, as given in Table~\ref{gradpn}. The open cluster 
results are shown as filled circles for the data by Friel et al. 
(\cite{friel02}), and as empty circles for the objects in the sample 
by Chen et al. (\cite{chen}), adopting the same criterium for the error 
bars. The [Fe/H] gradient from cepheid variables is shown as a filled 
star. Also included are the following young objects: OB stars in 
associations etc. from Daflon \& Cunha (\cite{daflon}, thick cross) and 
HII regions, as discussed in section~5 (x sign). The dotted vertical 
line shows the adopted age of the galactic disk, $t_G = 13.6$ Gyr, 
and the dashed line represents the time evolution of the [Fe/H] 
gradient according to the theoretical models of Hou et al. (\cite{hou}),
which are based on an inside-out scenario for the formation of the 
disk using metallicity--dependent yields.

It can be seen from Fig.~\ref{vargrad} that the open cluster gradients 
agree very well with the gradients derived from planetary nebulae, in 
the sense that both indicate a time flattening of the gradients during
the lifetimes of these objects, which is taken here as up to 8 Gyr. 
This conclusion holds in spite of the relatively large uncertainties
involved in the age determinations, especially in the case of the 
planetary nebula progenitor stars. Although the division of the PN and
open clusters into different age groups is arbitrary, it can be seen 
from Figs.~\ref{gradohv}, \ref{gradshv} and \ref{frielchen} that 
the time variation of the gradients is not sensitive to the particular 
groups chosen, as long as the age groups contain a reasonably large 
fraction of the total sample. 

The main uncertainty in the time variation of the gradients, as 
seen in Fig.~\ref{vargrad} refers to the youngest open clusters,
since, as mentioned, these objects are preferentially concentrated
in the inner Galaxy, so that the obtained correlation is not as
accurate as in the remaining cases. This may partially explain the
very low gradient found in Group I of both the Friel et al. 
(\cite{friel02}) and Chen et al. (\cite{chen}) samples.
Therefore, the gradient derived from the cepheid variable stars
plays a very  important role for ages under 1 Gyr, since it is well 
defined, as we have mentioned in section~4. As expected, this
gradient is steeper than those of Group I open clusters, but
it still clearly suggests some flattening in comparison with
the older objects in Fig.~\ref{vargrad}. Also in the young object
bracket, it is interesting to notice that both the OB stars of
Daflon \& Cunha (\cite{daflon}) and the HII regions discussed in
section~5 also show a very good agreement with the scenario 
displayed by the remaining objects, and are perfectly consistent
with the observed flattening during the last approximately 8 Gyr.

It is not our purpose here to discuss the effects of the present
results on existing models of the chemical evolution of the 
Galaxy, rather our aim is to provide observational constraints 
to these models. As an illustration, however, the observational 
data discussed here also show a good agreement with theoretical 
models by Hou et al. (\cite{hou}), as indicated by the dashed 
curve of Fig.~\ref{vargrad}. Naturally, our results refer to 
the last approximately 8 Gyr, which is the limit of the oldest 
age bracket for which we could derive some information. 
Nothing can then be said regarding the first 5 to 6 Gyr 
of the Galaxy lifetime, assuming a total age of $t_G = 13.6\ $Gyr,
so that our results do not contradict models that start with
a null gradient, which may then build up in a relatively short time.

The flattening rate of the [Fe/H] gradient is still uncertain, in 
view of the considerable error bars shown in Fig.~\ref{vargrad},
but from a simple analysis of all objects considered here,
an average rate of about 0.005 to 0.010 dex kpc$^{-1}$ Gyr$^{-1}$ 
can be obtained. This is similar to the earlier results by Maciel 
et al. (\cite{mcu2003}) and Chen et al. (\cite{chen}) for planetary 
nebulae and open clusters, respectively. Although this value is 
clearly uncertain, it sets the order of magnitude for the last
8 Gyr approximately, or from the time when the Galaxy was 5 to 6 Gyr
old to the present time.

\begin{acknowledgements}
      We thank P. A. L. Ferreira, P. S. Ribeiro and M. M. M. Uchida for
      some helpful discussions. This work was partly supported by  
      FAPESP, CNPq and CAPES.  Observations at ESO/Chile were possible 
      through the FAPESP grant 98/10138-8.
      
\end{acknowledgements}


\end{document}